\documentclass[aps,prl,superscriptaddress,reprint]{revtex4-2}
\usepackage{centernot}
\usepackage{graphicx}
\usepackage{amsmath}
\usepackage{amsbsy}
\usepackage{times}
\usepackage{amssymb}
\usepackage{mathrsfs}
\usepackage{chemarr}
\usepackage{xcolor}
\usepackage{url}
\usepackage{upgreek}
\usepackage{version}
\usepackage{ulem}
\usepackage[pdftex,colorlinks=true,pdfstartview=FitV,linkcolor= linkcolor,citecolor= linkcolor,urlcolor= linkcolor,hyperindex=true,hyperfigures=false]{hyperref}
\definecolor{linkcolor}{rgb}{0,0,0.6}
\usepackage{lipsum}
\usepackage{tikz}
\usepackage{accents}

\newcommand\br{{\bf r}}
\newcommand\bq{{\bf q}}

\newcommand\tu{{\tilde u}}

\newcommand\tes{{\tilde \ell^{\rm s}}}

\newcommand\rmH{{\rm H}}

\newcommand\rmd{{\rm d}}

\newcommand\mF{{\mathcal{F}}}
\newcommand\mE{{\mathcal{E}}}
\newcommand\mW{{\mathcal{W}}}

\newcommand\xic{{\xi^{\rm c}}}

\definecolor{redUP}{RGB}{138,21,56}

\graphicspath{{./Image/}}

\definecolor{gianmarco}{HTML}{ed2867}

\begin{document}

\title{Low-Pass Filtering of Active Turbulent Flows to Liquid Substrates}
\date{\today}

\author{Gianmarco Spera}
\email{gianmarco.spera@physics.ox.ac.uk}
\affiliation{Rudolf Peierls Centre for Theoretical Physics, University of Oxford, Oxford OX1 3PU, United Kingdom}
\author{Julia M. Yeomans}
\affiliation{Rudolf Peierls Centre for Theoretical Physics, University of Oxford, Oxford OX1 3PU, United Kingdom}\author{Sumesh P. Thampi}
\affiliation{Department of Chemical Engineering, Indian Institute of Technology, Madras, Chennai, India 600036}
\affiliation{Rudolf Peierls Centre for Theoretical Physics, University of Oxford, Oxford OX1 3PU, United Kingdom}

\begin{abstract}
To study the impact of active systems on their surroundings, we introduce a model that couples an active nematic fluid to an isotropic substrate fluid via friction. We numerically show that as the active layer develops turbulence, the substrate inherits the chaotic behaviour, exhibiting a novel form of turbulence driven by locally generated stochastic forcing from the active layer. In particular, the short-length-scale flow structures in the active layer are filtered out, so the system behaves as a de facto low-pass filter. We derive analytically the transfer function between the two layers and use it to predict the large-$q$ decay of the substrate energy spectrum, and to investigate how tensorial quantities, such as the strain rate and the active stresses, are transmitted between the active layer and the substrate. Our analysis agrees with recent experiments measuring velocity-velocity correlations in mixtures of active and passive microtubules, and it may have implications for traction force microscopy measurements in cellular layers.
\end{abstract}

\maketitle

Across all length scales, the non-equilibrium nature of active matter often manifests as spontaneous mechanical motion \cite{marchetti2013hydrodynamics,bechinger2016active,chate2020dry,o2022time}. A striking example is {\it active turbulence}~\cite{doostmohammadi2018active,alert2022active},  a chaotic flow state with strong vorticity, observed in reconstituted suspensions of cytoskeletal filaments and molecular motors \cite{sanchez2012spontaneous}, bacterial and microswimmer suspensions \cite{aranson2022bacterial}, cellular monolayer~\cite{lin2021energetics}, driven colloids \cite{mecke2023simultaneous} and granular rods \cite{narayan2007long}. 
Unlike externally driven systems, the dynamics of active matter is strongly influenced by boundaries and confinement \cite{bechinger2016active, liu2021viscoelastic,hardouin2022active,shankar2022topological,lacroix2024emergence,nishiyama2024closed,ardavseva2025beyond}, cases well studied in the literature. Conversely, the ways in which active systems drive and modify the dynamics of their surroundings remains comparatively unexplored. 

Emblematic experiments demonstrating active turbulence in suspensions of microtubule bundles powered by kinesin molecular motors are performed at fluid–fluid interfaces, with substrate liquids present on both sides of the interface~\cite{sanchez2012spontaneous,martinez2019selection,bantysh2024first}. The depth and properties of these fluids affect the turbulent motion~\cite{martinez2021scaling,thijssen2021submersed}. Similarly, epithelial cells display strikingly different correlations in their behaviour when seeded on substrates with different rheological properties~\cite{lo2000cell,angelini2010cell,murrell2011substrate,zheng2017epithelial,oakes2018lamellipodium,vazquez2022effect,charbonier2025substrate,yang2025dynamic}, showing sharp sensitivity to spatial variations~\cite{suh2024cadherin}. To capture the emergence of complex flow structures and order in active liquids, theoretical models have incorporated viscoelastic boundaries~\cite{mori2023viscoelastic}, spatially dependent activity~\cite{thijssen2021submersed}, and deformable~\cite{venkatesh2025nematic} or anisotropic substrates~\cite{pearce2019activity,parmar2025proliferating}. Yet in both experiments and theory, the influence of the neighbouring active material on substrate dynamics is often neglected. 
Indeed, active materials also offer potential for small-scale energy extraction~\cite{di2010bacterial,kaiser2014transport,thampi2016active} and fluid mixing~\cite{tayar2023controlling}, where understanding momentum transfer to an adjacent medium is essential.

In this \textit{Article} we use numerical and analytical approaches to study the dynamics of a liquid substrate driven by an active turbulent flow field, and demonstrate a low-pass (high frequency) filtering phenomenon mediated by frictional coupling between the two media. 
We show that the substrate inherits turbulent patterns from the active fluid but selectively filters out short length scales. We further show how stresses in the active layer are transmitted to the substrate and misalign with the substrate strain rate. 

The coupled active–passive system explored here constitutes a distinct form of turbulence. Unlike high–Reynolds-number or polymeric (viscoelastic) turbulence, the substrate turbulence emerges within an isotropic fluid, driven by locally generated, stochastic forcing.

\begin{figure}[t!]
  \begin{center}
    \begin{tikzpicture}
      \path (0,0) node {\includegraphics{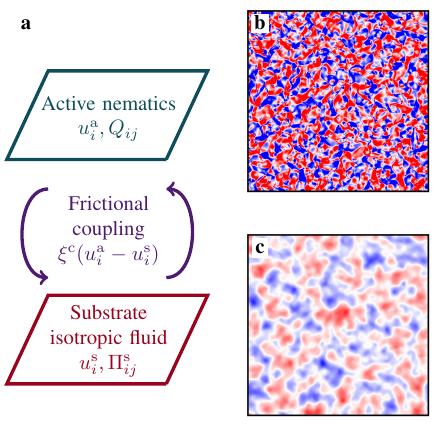}};

    \end{tikzpicture}
  \end{center}
  \caption{{\bf (a)} Schematic of the model: an active nematic layer, with velocity $u_i^{\rm a}$ and nematic order parameter $Q_{ij}$, is coupled via friction to a substrate layer, with velocity $u_i^{\rm s}$ and stress tensor $\Pi^{\rm s}_{ij}$. The magnitude of the coupling is set by the coefficient $\xi^{\rm c}$. {\bf (b-c)} Snapshots of typical flow fields in the active and the substrate layers; color represents the vorticity field $\omega^{\rm a|s} = (  \partial_y u_x^{\rm a|s} - \partial_x u_y^{\rm a|s} )/2$. Blue (red) regions correspond to counterclockwise (clockwise) rotation. The substrate inherits the turbulent flows from the active layer but  filters out the short length scales of the active turbulence.}
  \label{Fig:fig1}
\end{figure}

\begin{figure*}[ht!]
  \begin{center}
    \includegraphics{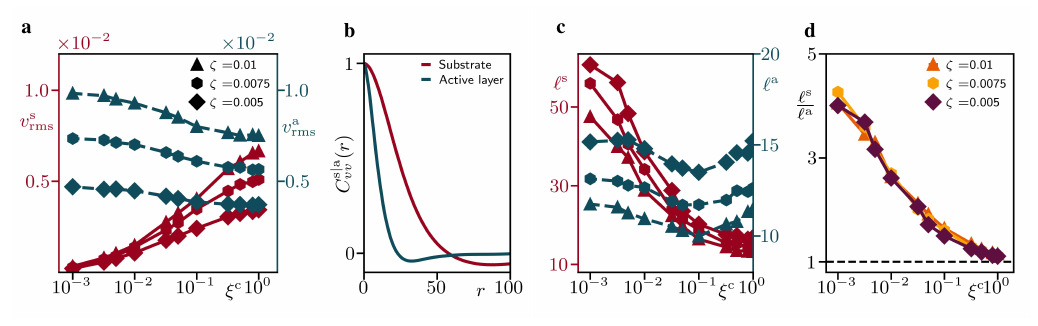}
  \end{center}
  
  \caption{
  {\bf (a)} Root-mean-squared velocity $v_{\rm rms}^{\rm s|a}$ in the substrate and active layers as the friction coefficient $\xic$ is varied for three different values of the activity $\zeta$. $v^{\rm s}_{\rm rms}$ increases as the friction coupling increases, but $v^{\rm a}_{\rm rms}$ decreases due to increased dissipation. For large $\xic$, $v^{\rm s}_{\rm rms} \approx v^{\rm a}_{\rm rms}$, the effective friction experienced by the active layer vanishes, and the substrate has the same flow structure as the active layer.  {\bf (b)} Example of velocity-velocity correlation functions in the substrate (red) and in the active layer (blue), $C_{vv}^{\rm s}$ and $C_{vv}^{\rm a}$ respectively, for $\zeta = 0.01$ and $\xic=0.01$. 
  $C_{vv}^{\rm s}$ decays over a longer length scale than $C_{vv}^{\rm a}$.
  {\bf (c)} Substrate and active length scales $\ell^{\rm s|a}$ respectively, extracted from $C_{vv}^{\rm s|a}(r)$ and
    {\bf (d)} the ratio $\ell^{\rm s}/\ell^{\rm a}$ as $\xic$ is varied for different activity coefficients. $\ell^{\rm s}/\ell^{\rm a} > 1$ and, as the coupling coefficient increases, decreases to unity. This curve quantifies the low-pass filtering that has occurred in the substrate turbulence, and the role of friction in this process. 
  Correlation lengths $\ell^{\rm s|a}$ are defined by the conditions $C_{vv}^{\rm s|a}(\ell^{\rm s|a}) = 1/e$. Note that in (c) we use different scales for $\ell^{\rm s}$ and $\ell^{\rm a}$. Error bars are smaller than symbols.
  Similar behaviour to {\bf (c)}-{\bf (d)} is observed for time velocity correlations, as shown in Note S3~\cite{supp}. 
  }  
  \label{Fig:fig2}
\end{figure*}

{\it Model.} The system comprises a two-dimensional active fluid interacting with a two-dimensional liquid substrate through frictional coupling, as shown in Fig.~\ref{Fig:fig1}a. We will use the superscripts ${\rm a}$ and ${\rm s}$ to refer to the active and substrate layers respectively. 
The active layer is modeled as an active nematic fluid~\cite{doostmohammadi2018active}, which has proved to be a suitable continuum description for a variety of active systems~\cite{dell2018growing,saw2018biological,maroudas2021topological,balasubramaniam2022active}.  
Nematic ordering in the active layer is described by a tensorial order parameter $Q_{ij} = 2S[n_in_j-\delta_{ij}/2]$, where $S$ quantifies the magnitude of the nematic order and $n_i$ is the director field. 
(Note that we use the Cartesian index notation for vectors and tensors and assume Einstein summation convention for repeated indices.)  
The nematic tensor evolves as 
\begin{equation}\label{eq:dynamics-Q}
  D_t Q_{ij} - \mW_{ij} = \Gamma \rmH_{ij} \;, 
\end{equation}
where $D_t = \partial_t + u_k^{\rm a} \partial_k$ is the material derivative, and $u_k^{\rm a}$ is the velocity field in the active layer. 
The generalized advection term,
$\mW_{ij} = \left(\lambda E^{\rm a}_{ik} + \omega^{\rm a}_{ik}\right) \tilde Q_{kj} + \tilde Q_{ik} (\lambda E^{\rm a}_{kj} - \omega^{\rm a}_{kj}) - 2 \lambda \tilde Q_{ij} \left(Q_{kl}\partial_k u^{\rm a}_l\right)$,
encodes advection of the order parameter due to velocity gradients. Here $\tilde Q_{ij} \equiv Q_{ij} + \delta_{ij}/2$, $E^{\rm a}_{ij} = (\partial_i u^{\rm a}_j + \partial_j u^{\rm a}_i)/2$ is the strain rate tensor, 
$\omega^{\rm a}_{ij} = (\partial_j u^{\rm a}_i - \partial_i u^{\rm a}_j)/2$ is the vorticity tensor, and $\lambda$ is the flow alignment parameter. 
The rotational diffusivity $\Gamma$ controls the relaxation to the minimum of the Landau-de Gennes free energy~\cite{de1993physics}, $\mF_{\rm LDG} = \int \rmd \br f_{\rm LDG}(\br)$, where $f_{\rm LDG} = \frac{A}{2} \left[ S_0^2 - \frac{1}{2} \left( Q_{ij}^2\right) \right]^2 +  \frac{K}{2} (\partial_k Q_{ij})^2$, through the molecular field $\rmH_{ij} = - \left[ { \delta \mF_{\rm LDG} }/{\delta Q_{ij}} - ({\delta_{ij}}/{2}) {\rm Tr} \left( { \delta \mF_{\rm LDG} }/{\delta Q_{ij}}\right) \right]$. Here $S_0$ is the nematic order at equilibrium, $K$ is the elastic constant, and $A$ sets the energy scale for the bulk free energy. 

The velocity field $u_i^{\rm a}$ obeys the continuity and the incompressible Navier-Stokes (NS) equations:
\begin{equation}\label{eq:ns-active}
 \partial_k u_k^{\rm a} = 0\;, \quad \rho^{\rm a} D_t u^{\rm a}_i = \partial_j \Pi^{\rm a}_{ij} - \xi^{\rm c}( u^{\rm a}_i - u^{\rm s}_i ) \;, 
\end{equation} 
where $\rho^{\rm a}$ is the density in the active layer. The total stress in the active nematic layer $\Pi^{\rm a}_{ij}$ can be decomposed as $\Pi^{\rm a}_{ij} = \Pi_{ij}^{\rm pass} + \Pi_{ij}^{\rm visc} + \Pi_{ij}^{\rm act}$ where $\Pi_{ij}^{\rm pass} = -P^{\rm a}\delta_{ij} + 2 \lambda \tilde Q_{ij}\left( Q_{kl} \rmH_{kl} \right)  - \lambda \rmH_{ik} \tilde Q_{kj}  - \lambda \tilde Q_{ik} \rmH_{kj}   - \left( \partial_i Q_{kl} \right){\delta \mF_{\rm LDG}}/{\delta (\partial_j  Q_{kl})  } + Q_{ik} \rmH_{kj} - \rmH_{ik} Q_{kj} $ is a passive stress that generates backflow with $P^{\rm a}$ being the pressure in the active layer; 
$\Pi_{ij}^{\rm visc} = 2 \eta^{\rm a} E^{\rm a}_{ij}$ is the viscous stress, with $\eta^{\rm a}$ the viscosity of the active fluid; 
$\Pi_{ij}^{\rm act}= - \zeta Q_{ij}$ the active stress, where a positive (negative) activity coefficient $\zeta$ corresponds to an extensile (contractile) system. 

The coupling of the active layer to the substrate is introduced via  friction $- \xic( u^{\rm a}_i - u^{\rm s}_i )$ in Eq.~\eqref{eq:ns-active}, where  $u_i^{\rm s}$ is the velocity of the substrate layer. The intensity of the coupling is controlled by the coefficient $\xic$. The frictional force vanishes if the relative velocity between the two layers is locally zero. 
The velocity field in the substrate, $u_i^{\rm s}$, also satisfies the continuity and the incompressible NS equations: 
\begin{equation}\label{eq:ns-substrate}
  \partial_k u_k^{\rm s} = 0\;, \quad \rho^{\rm s} D_t u_i^{\rm s} = \partial_j \Pi^{\rm s}_{ij} + \xi^{\rm c}( u^{\rm a}_i - u^{\rm s}_i ) \;,
\end{equation}
where $\rho^{\rm s}$ is the density and $\Pi^{\rm s}_{ij}$ is the stress tensor capturing the rheological response of the substrate fluid. 

In the numerical results demonstrating low-pass filtering presented here, we focus on the case of a Newtonian substrate fluid, \textit{i.e.,}  $\Pi^{\rm s}_{ij} = -P^{\rm s} \delta_{ij} + 2\eta^{\rm s} E_{ij}^{\rm s}$, where $P^{\rm s}$ the pressure, $\eta^{\rm s}$ is the viscosity, and  $E_{ij}^{\rm s}$ is the strain rate tensor  in the substrate. However, our analytical arguments suggest that the conclusions are robust to different choices of $\Pi^{\rm s}_{ij}$. As examples, cases of viscoelastic fluid substrates are discussed in Note S2~\cite{supp}.

Equations~\eqref{eq:dynamics-Q}--\eqref{eq:ns-substrate} are solved using a hybrid lattice Boltzmann (LB) method~\cite{marenduzzo2007steady,thampi2014vorticity,kruger2017lattice}: Eqs.~\eqref{eq:ns-active}-\eqref{eq:ns-substrate} are solved using LB while Eq.~\eqref{eq:dynamics-Q} is solved using finite-difference methods.
We use periodic boundary conditions, initialize the system in a no flow and disordered nematic state, and then run simulations for $T=10^5$ LB time steps.
Parameters common to all results are: $L_x=L_y=400$, $\Gamma = 0.1$, $A=0.1$, $K=0.01$, $S_0=1$, $\lambda=0.3$, $\eta^{\rm a}=\eta^{\rm s} = 2/3$, and $\rho^{\rm a} =\rho^{\rm s} = 20$.

{\it Active turbulence generates substrate turbulent flows.}
Figure~\ref{Fig:fig1}(b-c) shows snapshots of the turbulent velocity field developed in the active and the substrate layers (see also Movie S1 \cite{supp}). Due to the action of the active stress the well known hydrodynamic instabilities and the state of active turbulence develop in the active layer~\cite{doostmohammadi2018active}. The frictional coupling, which is proportional to the relative velocity between the layers, transfers part of the momentum generated in the active layer to the substrate. Consequently, the substrate layer also develops turbulent flows.

The momentum transfer to the substrate can be quantified by measuring the root-mean squared velocity $v_{\rm rms}^{\rm s|a} = [\langle ( u_i^{\rm s|a} )^2 \rangle]^{1/2}$ in the two layers.
As shown in Fig.~\ref{Fig:fig2}a, $v^{\rm s}_{\rm rms}$ increases monotonically with increase in the friction coupling coefficient $\xic$ due to more effective momentum injection from the active layer. But increased frictional coupling implies increased dissipation in the active layer,  so the velocity in the active layer $v^{\rm a}_{\rm rms}$ decreases with increase in $\xic$. 
With increase in $\xic$, $v^{\rm a}_{\rm rms} \approx v^{\rm s}_{\rm rms}$ as the relative velocity between the two layers decreases and they   behave as a single fluid.
Further, Fig.~\ref{Fig:fig2}a  shows the change in $v_{\rm rms}$ in the two layers for different activities. As expected, an increase in the activity of the active nematic layer increases $v^{\rm a}_{\rm rms}$, and hence  $v^{\rm s}_{\rm rms}$. However, the way in which the velocity in the active layer and the substrate layer vary with activity depends on the coupling coefficient $\xic$. This difference arises because substrate turbulence does not exactly mirror the features of active turbulence.

{\it Substrate filters out short-length scales.}
To quantify the differences in the flow structures in the substrate and active layers, we measure the velocity-velocity spatial correlations in both layers, $C^{\rm s|a}_{vv}(\br) = \langle u_k^{\rm s|a}(\br+\br_0) u_k^{\rm s|a}(\br_0) \rangle_{\br_0, t}$, where the average is taken over time and initial positions $\br_0$. Since the system is isotropic $C^{\rm s|a}_{vv}$ is only a function of the distance $r=|\br-\br_0|$. Figure~\ref{Fig:fig2}b shows that $C_{vv}^{\rm s}$ extends to  larger $r$ than $C_{vv}^{\rm a}$ indicating that the smaller length scales are suppressed by the momentum transfer between the layers.

To further quantify this filtering effect we calculate the velocity correlation lengths in the substrate and the active layer, $\ell^{\rm s}$ and $\ell^{\rm a}$ respectively, from the corresponding correlation functions $C_{vv}^{\rm s|a}(r)$. Figures~\ref{Fig:fig2}c-d show the variation of these length scales as $\xic$ is varied. For small $\xic$, $\ell^{\rm s}$ can reach up to four times $\ell^{\rm a}$. As $\xic$ increases, $\ell^{\rm a}$ decreases as the active layer experiences increased frictional screening of the hydrodynamics. However, at larger values of $\xic$, $\ell^{\rm a}$ starts to increase again. This is because the relative velocity between the two layers decreases (Fig.~\ref{Fig:fig2}a) resulting in less dissipation and less hydrodynamic screening in the active layer. For large frictional coupling the active-substrate fluid system behaves as a single layer.

\begin{figure}[t!]
  \begin{center}
    \begin{tikzpicture}
      
      \path (0,0) node {\includegraphics[]{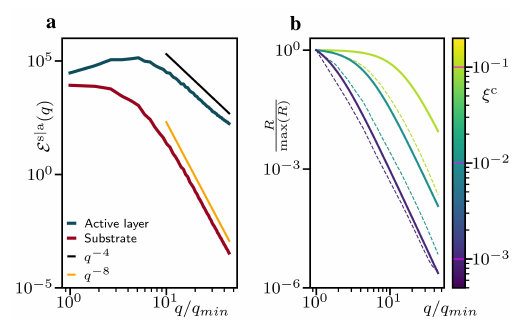}};
    \end{tikzpicture}
  \end{center}
  \caption{{\bf (a)} Energy spectra in the substrate layer $\mE^{\rm s}(q)$ and in the active layer $\mE^{\rm a}(q)$ for $\zeta =0.01$ and $\xic=0.01$. For large $q$, $\mE^{\rm s}(q)$ decays as $q^{-8}$ as opposed to $\mE^{\rm a}(q)$ which decays as $q^{-4}$, consistent with the predictions of the kinetic energy transfer function, Eq.~\eqref{eq:energy-spectra}. {\bf (b)} Spectral energy ratio $R \equiv \mE^{\rm s}(q)/\mE^{\rm a}(q)$ as the friction coupling is varied (for $\zeta = 0.01$); continuous and dashed lines are Eq.~\eqref{eq:energy-spectra} and from simulations respectively. Axes in {\bf (b)}  are normalized to make the comparisons easier.
  }
  \label{Fig:fig3}
\end{figure}

{\it Transfer functions for substrate turbulence.}
To substantiate the observed phenomenology of substrate turbulence, we now derive the transfer function between the velocity fields of the substrate and that of the active layer. This analytical result will also enable us to relate the kinetic energy spectrum and the velocity correlation function of the flow fields in the two layers.

To proceed, we consider the low Reynolds number limit of Eq.~\eqref{eq:ns-substrate}, 
to obtain the balance between fluid stress and frictional coupling force for the substrate flows: 
\begin{equation}\label{eq:simplified-nstv}
  0 = -\partial_iP^{\rm s} + 2\eta^{\rm s} \partial_j E_{ij}^{\rm s} + \xic ( u_i^{\rm a} - u_i^{\rm s}) \;.
\end{equation}
Fourier transforming  Eq.~\eqref{eq:simplified-nstv}, projecting out pressure, and using the incompressibility of the substrate and active fluids, 
(see Note S1), Eq.~\eqref{eq:simplified-nstv} reduces to a relation between the velocity fields in the two layers,
\begin{equation}\label{eq:transfer-velocity}
  \tu_i^{\rm s}(\mathbf{q}) = T({q}) {\tu_i^{\rm a}}(\mathbf{q})\;, \quad T({q}) = {[ 1+ (\tilde{\ell}^{\rm s})^2 q^2 ]^{-1} }  \;,
\end{equation}
where $\tu_i^{\rm s|a}$ is the Fourier transform of the velocity in the substrate and active layers, $T({q})$ is the {\it transfer function} connecting them, and $q = |\mathbf{q}|$. The length scale $\tilde{\ell}^s = \sqrt{\eta^{\rm s}/\xic}$ is the  screening length for the substrate hydrodynamics.
Equation~\eqref{eq:transfer-velocity} shows that the velocity field of the substrate turbulence is modulated relative to that of the active turbulence via the transfer function $T(q)$. This relation gives rise to the following two results that highlight its implications.

First, we derive the relation between the energy spectra of the two layers. Since, the spectral energy per unit mass is defined as $\mE^{\rm s|a}(q) = \frac{1}{2}|\tilde{u}_i^{\rm s|a}|^2$, 
a quantity determined solely by the respective velocity fields, the energy spectra in the substrate and the active layers are related by:
\begin{equation}\label{eq:energy-spectra}
  \mE^{\rm s}(q) = T^2(q) \mE^{\rm a}(q) = \frac{\mE^{\rm a}(q)}{[1+ (\tilde\ell^{\rm s} q)^2 ]^2}\;.
\end{equation}
Eq.~\eqref{eq:energy-spectra} shows that when the active layer injects energy at a wave number ${q}$, the kinetic energy transfer to the substrate
is filtered by $T^2(q)$, with an associated screening mode $q^{\rm s} = 2\pi/\tilde{\ell}^{\rm s}$. For $q \ll q^{\rm s}$, $T(q) \to 1$, implying no filtering during the momentum transfer, so the substrate inherits the large-scale behavior of the active layer. In contrast, for $q > q^{\rm s}$, energy transfer is increasingly suppressed and thus, short-wavelength structures from the active layer are filtered out. 

Since the wave number associated with hydrodynamic screening in the substrate layer, $q^{\rm s}\propto \sqrt{\xi^{\rm c}}$, an increase in $\xi^{\rm c}$ raises $q^{\rm s}$, which extends the range of $q$ over which $T(q)\simeq 1$. Therefore, with increase in $\xic$, substrate and active-layer flow characteristics become more similar ( $v^{\rm s}_{\rm rms} \approx v^{\rm a}_{\rm rms}$,  $\ell^{\rm s} \approx \ell^{\rm a}$) as evident from the plots in Fig.~\ref{Fig:fig2}.

Figure~\ref{Fig:fig3}a presents an example (corresponding to the parameters of Fig.~\ref{Fig:fig2}b) of the kinetic energy spectra of the two layers, $\mE^{\rm s|a}(q)$. The active layer exhibits the universal $q^{-4}$ decay at large $q$~\cite{alert2020universal,alert2022active}, whereas the substrate turbulence shows a steeper $q^{-8}$ decay. The resulting $\sim q^{4}$ difference between the two is in agreement with Eq.~\eqref{eq:energy-spectra}, the kinetic energy transfer function. To further test the validity of Eq.~\eqref{eq:energy-spectra}, we define the spectral energy ratio $R \equiv \mE^{\rm s}(q)/\mE^{\rm a}(q)$ and plot this in Fig.~\ref{Fig:fig3}b for different values of the friction coupling coefficient, at fixed activity $\zeta$. The comparison between measurements from the simulations (dashed lines) and the predictions of Eq.~\eqref{eq:energy-spectra} (continuous lines) improves as the frictional coupling weakens (smaller $\xic$). Indeed, as $\xic\to0$, $T(q) \sim 1/q^2$ and $R$ is thus dominated by the low-pass filtering effect. As $\xic$ increases, the contribution $q/q^{\rm s}$ is of the same order of the neglected contribution and the prediction worsen. 
In Note S1, we show how decreasing the Reynolds number of the substrate extends the range of validity of Eq.~\eqref{eq:energy-spectra}, and how, in all cases, the characteristic $q^{-4}$ decay of $R$ is clearly observed (see Fig.~S1).

We next relate the velocity correlation function in the substrate to that in the active layer. Since the velocity correlation function is the inverse Fourier transform of the corresponding spectral energy, $C_{vv}^{\rm s|a}(r) = \mF^{-1}(|{\tu}^{\rm s|a}_{i}(q_j)|^2)$, Eq.~\eqref{eq:transfer-velocity} allows us to write the velocity-velocity correlations in the substrate as
\begin{equation}\label{eq:vv-substrate}
  C_{vv}^{\rm s}(r) =
  \mathcal{F}^{-1} \left\{ T^2(q) \mF[C_{vv}^{\rm a}(r)]\right\} \;,
\end{equation}
where $ \mF[C_{vv}^{\rm a}(r)] $ is the direct Fourier transform of $C_{vv}^{\rm a}(r)$. Eq.~\eqref{eq:vv-substrate} illustrates that $T^2(q)$ modulates the velocity correlations between the two layers (Fig.~\ref{Fig:fig2}).  In the simple case where an active layer has exponential velocity correlations $C_{vv}^{\rm a}(r) \sim e^{-r/\ell^a}$ it is possible to make analytic progress to reach an expression that can then be calculated numerically (see Note S1). This confirms that the  substrate length scale is indeed larger than $\ell^a$ for all parameter values.

The arguments presented in this section show that the features inherited by the substrate turbulence crucially depend upon the strength of frictional coupling between the two layers, but not the specific constitutive relation chosen for the substrate. Hence we expect that the low-pass filtering of active turbulent flows also occurs when the substrate is not Newtonian. This is demonstrated using both the Kelvin-Voigt and the Maxwell model for the viscoelastic substrate, see Note S2. 

Finally, we mention that, similar to the phenomena of low pass filtering of small length scales, low pass filtering of short time scales also occurs between the active and substrate turbulent fields  (Note S3).

\begin{figure}[t!]
    \centering
    \begin{tikzpicture}
    \def\x{2}
    \def\y{4}
    \def\dx{3}
    \def\dy{3}

    \path (0,0) node {\includegraphics{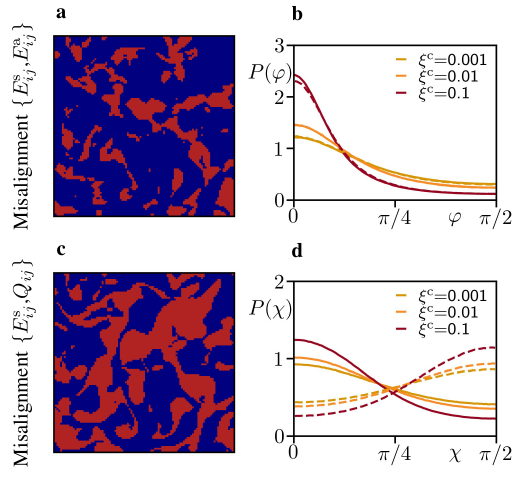}};

    \end{tikzpicture}
    \caption{{\bf (a-c)} Simulation snapshots (with parameters as in Fig.~\ref{Fig:fig1}) colored according to $\varphi$ and $\chi$, defined as the angles between the director fields of the pairs $\{E_{ij}^{\rm s}, E_{ij}^{\rm a}\}$ and $\{E_{ij}^{\rm s}, Q_{ij}\}$ respectively. Regions in which $\varphi,\chi<\pi/4$ are colored blue and regions with $\varphi,\chi\ge\pi/4$ are colored red.  {\bf (b-d)} Probability distribution of $\varphi$ and $\chi$ as the friction coupling $\xic$ is varied for extensile (continuous lines) and contractile (dashed lines) activity $|\zeta|=0.01$. }
    \label{fig:fig4}
\end{figure}
{\it Strain rate alignment in substrate turbulence.} 
So far, we have analyzed scalar and vector measures, namely the energy spectrum and the velocity field of the substrate turbulence, and compared them to their counterparts in the active layer. Here, we compare the characteristics of a tensorial field, the strain rate in the substrate $E^{\rm s}_{ij}$, with two tensorial fields in the active layer, the strain rate $E^{\rm a}_{ij}$ and the nematic order parameter $Q_{ij}$. These variables encode information about the viscous flows and active stress (as $\Pi_{ij}^{\rm act} = - \zeta Q_{ij}$) respectively. 

Fig.~\ref{fig:fig4}a-c, show typical simulation snapshots of the angles $\varphi$ and $\chi$ between the director fields (normalised largest eigenvectors) of the pairs $\{E_{ij}^{\rm s}, E_{ij}^{\rm a}\}$ and $\{E_{ij}^{\rm s}, Q_{ij}\}$. 
In the plots blue denotes that the angle is $<\pi/4$, and red an angle $>\pi/4$. The results show clear regions of misalignment ($>\pi/4$) between the tensor fields in the active layer and the passive substrate, reminiscent of those reported in epithelial cell layers \cite{nejad2024stress}. The probability distribution functions of the angles, 
$P(\varphi)$ and $P(\chi)$, are shown in Fig.~\ref{fig:fig4}b-d for different friction coefficients and for both extensile and contractile activity (see also Movies S2-3). Note that $P(\varphi)$ is independent of activity $\zeta$ while $P(\chi)$ is not.

It is possible to derive an analytical expression for the misalignment angle $\varphi$ between the strain-rate tensors in real space, see Note S4. 
Indeed, Eq.~\eqref{eq:transfer-velocity} also relates the strain rates via $E_{ij}^{\rm s}(\bq) = T(q)E_{ij}^{\rm a}(\bq)$, which then allows to rewrite them in real space as
\begin{equation}\label{eq:real-space-strain}
    E_{ij}^{\rm s}(\br) = \int \rmd \br' \; K_0\left(\frac{|\br - \br'|}{\tes}\right)E^{\rm a}_{ij}(\br') \;,
\end{equation}
where $K_0$ is the modified Bessel function of the second kind. A straightforward consequence is that, within our approximations, the strain rates are aligned in Fourier space as $T(q)$ affects the moduli but not the directors. The real-space misalignment is thus a result of the non-local convolution in Eq.~\eqref{eq:real-space-strain}.
Therefore, the angle $\varphi$ can be expressed locally as 
\begin{equation}\label{eq:convolution-strain-strain}
    \cos(2\varphi(\br)) = 
    \frac{E^{\rm a}_{ij}(\br)  \int \rmd \br'\; K_0\left(\frac{|\br - \br'|}{\tes}\right)E^{\rm a}_{ij}(\br') }{|E_{ij}^{\rm a} (\br)| |\int \rmd \br'\; K_0\left(\frac{|\br - \br'|}{\tes}\right)E^{\rm a}_{ij}(\br') | } \;, 
\end{equation}
where $|\cdot|$ is the norm $|A_{ij}| \equiv \sqrt{A_{ij} A_{ij}}$ that ensures tensor normalizations. The results observed in Fig.~\ref{fig:fig4}b can now be understood from Eq.~\eqref{eq:convolution-strain-strain}. As $\tilde \ell^{\rm s}\to 0$ ($\xic\to \infty$), $K_0(r/\tilde \ell^{\rm s}) \sim {\rm e}^{-r / \tes } / r^{1/2}$ suppresses spatial averaging and leads to $E^{\rm s}_{ij} \sim E^{\rm a}_{ij} $. 
Thus, $E_{ij}^{\rm s}$ and $E_{ij}^{\rm a}$ are more aligned and the distribution $P(\varphi)$ shifts towards $\varphi=0$. On the other hand, as $\tilde \ell^{\rm s}\to \infty$ ($\xic\to 0$), $K_0(r/\tilde \ell^{\rm s}) \sim - {\rm ln}(r /\tes)$ leads to stronger non-local effects in the averaging process and $P(\varphi)$ widens and is less peaked at zero. 
While a full determination of $P(\varphi)$ and $P(\chi)$ in real space is out of reach, in Note S4 we show how we can make analytical progress in predicting the sign of the spatial averages of $\varphi$ and $\chi$ under the assumption of fast relaxing norms. 
There, we explicitly show how $P(\chi)$ depends on the type of activity considered, contrary to $P(\varphi)$ see Fig.~\ref{fig:fig4}b-d.

{\it Summary and discussion.}
In this paper we studied the impact of active systems on their surroundings by studying an isotropic substrate liquid coupled to an active nematic layer. 
We showed that as the active layer develops turbulence, it dissipates momentum into the substrate so that the substrate fluid inherits the turbulent flows. However, the substrate turbulence develops over larger length scales than those in the active layer. To rationalise this process, we derived an analytical expression for the transfer function relating the 
velocity fields in the active and substrate layers. The transfer function showed that energy injection to the substrate is suppressed below a given characteristic length scale $\tilde \ell^{\rm s}$ leading to  low-pass filtering of the active flows in both space and time. 
We also studied how the characteristics of tensorial fields in the active layer are transferred to the substrate and demonstrated the emergence of misalignment between the strain rate field in the substrate and both the strain rate and active stress fields in the active layer. 

Recent experiments involving mixtures of active and passive microtubules have shown the low-pass filtering effect in length- and time-scales on measuring velocity-velocity correlations~\cite{berezney2024controlling}. 
Our analysis may also have implications for traction force microscopy measurements of stress in cellular layers, as recent studies suggest stress-cell shape misalignment in MDCK monolayers~\cite{nejad2024stress}. We show here that misalignment is a characteristic of the coupled system, and the extent of misalignment between various tensor fields depend upon the strength of coupling and the activity in the system. Therefore our results may be relevant to better understanding how measurements inferred via a substrate relate to the corresponding quantities in an active cellular layer. It will be interesting in future work to consider elastic substrates, most relevant to traction force microscopy experiments, or liquid crystal substrates~\cite{bantysh2024first}. 

Unlike externally driven systems that give rise to high Reynolds number or low Reynolds number polymeric (viscoelastic) turbulent flows, here the forcing is local and random, yet the substrate turbulence differs from active turbulence because the substrate fluid is Newtonian and follows a linear constitutive relation. Thus this may constitute a platform to investigate the properties of a different form of low Reynolds number turbulence or to rejuvenate interest in the application of renormalization group theory developed for randomly stirred fluids  \cite{forster1976long, forster1977large} to experimentally realizable examples.

\vspace{0.2cm}
{\it Acknowledgments.}
We thank Delia Cropper, Francesco Mori, Ioannis Hadjifrangiskou, Rahil Valani, and Vedad Dzanic for fruitful discussions.  GS and JMY acknowledge support from the ERC Advanced Grant ActBio (funded as UKRI Frontier Research Grant EP/Y033981/1). SPT thanks the Royal Society and the Wolfson Foundation for the Royal Society Wolfson Fellowship award and acknowledges the support of the Department of Science and Technology, India via the research grant CRG/2023/000169.

\nocite{morozov2015introduction}

\bibliography{biblio_clean.bib}
\end{document}